\documentclass[10pt]{article}

\usepackage{amsmath}
\usepackage{amssymb}

\usepackage{graphicx}

\usepackage{cite}

\usepackage[usenames, dvipsnames]{color}
\usepackage[normalem]{ulem}

\usepackage[labelfont=bf,labelsep=period,justification=raggedright]{caption}

\makeatletter
\renewcommand{\@biblabel}[1]{\quad#1.}
\makeatother

\date{}

\newcommand{\half}{\tfrac{1}{2}}

\newcommand{\sM}{\sigma_{\rm m}}
\newcommand{\sS}{\sigma_{\rm s}}
\newcommand{\eOpt}{e_{\rm opt}}
\newcommand{\eMax}{e_{\rm max}}
\newcommand{\sThres}{\tilde{s}}
\newcommand{\avUni}{\mu}
\newcommand{\avDyn}{\mu}
\newcommand{\avBin}{\mu}
\newcommand{\f}{\frac}
\newcommand{\bi}{\begin{itemize}}
\newcommand{\ei}{\end{itemize}}
\newcommand{\be}{\begin{equation}}
\newcommand{\ee}{\end{equation}}
\newcommand{\payoff}{F} 
\newcommand{\cost}{C}
\newcommand{\benefit}{B}
\newcommand{\costConst}{c}
\newcommand{\nT}{\ell}
\newcommand{\md}{{\mathrm d}}
\newcommand{\pa}{\partial}

\def\l{\left}
\def\r{\right}

\begin{document}

\begin{flushleft}
{\Large
\textbf{Environmental statistics and optimal regulation}
}
\\
David A. Sivak$^{\ast\dagger}$ and Matt Thomson$^{\ast\dagger}$
\\
Center for Systems and Synthetic Biology, University of California, San Francisco, California, USA
\\
$\ast$ E-mail: david.sivak@ucsf.edu, matthew.thomson@ucsf.edu
\\
$\dagger$ These authors contributed equally to this work.
\end{flushleft}

\section*{Abstract}
Any organism is embedded in an environment that changes over time. The timescale for and statistics of environmental change, the precision with which the organism can detect its environment, and the costs and benefits of particular protein expression levels all will affect the suitability of different strategies---such as constitutive expression or graded response---for regulating protein levels in response to environmental inputs. We propose a general framework---here specifically applied to the enzymatic regulation of metabolism in response to changing concentrations of a basic nutrient---to predict the optimal regulatory strategy given the statistics of fluctuations in the environment and measurement apparatus, respectively, and the costs associated with enzyme production. We use this framework to address three fundamental questions: (i) when a cell should prefer thresholding to a graded response; (ii) when there is a fitness advantage to implementing a Bayesian decision rule; and (iii) when retaining memory of the past provides a selective advantage. We specifically find that: (i) relative convexity of enzyme expression cost and benefit influences the fitness of thresholding or graded responses; (ii) intermediate levels of measurement uncertainty call for a sophisticated Bayesian decision rule; and (iii) in dynamic contexts, intermediate levels of uncertainty call for retaining memory of the past. Statistical properties of the environment, such as variability and correlation times, set optimal biochemical parameters, such as thresholds and decay rates in signaling pathways. Our framework provides a theoretical basis for interpreting molecular signal processing algorithms and a classification scheme that organizes known regulatory strategies and may help conceptualize heretofore unknown ones.

\section*{Author Summary}
All organisms live in environments that dynamically change in ways that are only partially predictable. The seasons, diurnal cycles, oceanic fluid dynamics, the progression of food through the human gut, all impose some predictability on common microbial ecosystems. Microbes are also at the whim of random processes (like thermal motion) that introduce uncertainty into environmental change.  Here, we develop a theoretical framework to analyze how cellular regulatory systems might balance this predictability and uncertainty to most effectively respond to a dynamic environment. We model a simple cellular goal: regulating a single enzyme to maximize the energy generated from a nutrient whose environmental concentration varies. In this context, optimal regulatory strategies are determined by an uncertainty ratio comparing cellular measurement noise and environmental variability. Intermediate levels of uncertainty call for sophisticated Bayesian decision rules, where selective advantage accrues to organisms that incorporate past experience in their inference of the current environmental state. When uncertainty is either high or low, optimal signal processing strategies are comparatively simple: constitutive expression or naive tracking, respectively. This work provides a theoretical basis for interpreting molecular signal processing algorithms and suggests that relative levels of environmental variability and cellular noise affect how microbes should process information.

\section*{Introduction}
Any organism is embedded in an environment that changes in ways that are typically outside the organism's control and stochastic, yet not entirely unpredictable. In response to such changing environmental conditions, organisms dynamically regulate the expression of their genomes to meet physiological demands~\cite{LopezMaury:2008}. For example, microorganisms implement circuits of signal transduction and regulation that collect information from the environment and modulate expression of metabolic enzymes to convert environmental nutrients into energy for functional goals such as protein production, cell growth, and division~\cite{Zaman:2008,Barkai:1997cd}.  

For environmental sensing and gene regulation, biomolecular circuits often employ complex information processing and control algorithms~\cite{Perkins:2009cg} that can be schematically classified into broad and qualitatively-distinct classes, including: insensitivity to environmental conditions, sensing changes and then responding, temporal averaging~\cite{Hersen:2008fe}, adaptation~\cite{Yi:2000}, stochastic switching~\cite{Kussell:2005dg}, or prediction of future changes on the basis of past conditions~\cite{Tagkopoulos:2008ct,Mitchell:2009el}. An important goal of systems biology is to catalog the molecular circuits~\cite{Lim:2013eo} and corresponding information processing algorithms~\cite{Nurse:2008en} used by a range of organisms and to understand how information processing algorithms are adapted to particular cellular tasks like metabolic regulation as well as to particular environmental niches~\cite{Perkins:2009cg}. 

Microorganisms occupy a diverse range of environmental niches, so that characteristic time scales of environmental change range over many orders of magnitude~\cite{Stocker:2012ea,FriasLopez:2008,Demir:2011}. Temporal correlations in environmental structure emerge through day and night cycles, seasons, weather patterns, timescales of host dynamics, and complex physical processes like fluid flow, turbulence, and diffusion~\cite{Vergassola:2007dj,Shraiman:2000er,Berg:1977bp,Rust:2011gq}. Intuitively, various architectures of sensing and control circuits will differ in their suitability across a range of environmental statistical patterns and dynamic time scales, but a rigorous connection is lacking. Put concretely, when does it make sense to ignore one's surroundings, to trust one's immediate senses, to do more complicated inference, or to remember the past?

Here, we develop a general decision-theoretic framework for deriving optimal regulatory algorithms for a model cellular task---the regulation of expression of a single enzyme in response to a time-varying environmental nutrient concentration~\cite{Dekel:2005ey,Kalisky:2007}---given the statistics of environmental fluctuations, measurement precision, and enzymatic expression costs. Whereas much research has focused on how to achieve particular regulatory functions, here we focus on the related question of how preferences for different regulatory strategies depend on stochastic characteristics of the cell and environment. The timescales for environmental change, the statistical properties of the environment, and the precision with which the organism can detect its environment all will affect the suitability of different regulatory strategies. We demonstrate how different regimes of these basic physical properties of the environment and cell demarcate common signal processing strategies. For example, with perfect nutrient sensors, it is optimal for the cell to simply respond to the measured concentration of a nutrient signal; as sensors become noisy, the optimal strategy switches to one of internalization through Bayesian priors of the statistics of environmental dynamics, which overcomes inherent physical limitations in measurement precision.

Previous studies have postulated a role for Bayesian decision rules in nutrient sensing and studied biochemical implementations of optimal Bayesian sensing strategies in a limited number of circumscribed environmental contexts~\cite{Libby:2007fw}. In our framework, Bayesian inference emerges as a natural consequence of maximizing enzymatic benefit, averaged over a probabilistic environment. Further, our theoretical framework enables analytical calculation of optimal enzymatic regulatory strategies over a large range of different environmental statistics.

\section*{Results/Discussion}

\subsection*{Model system: regulation of a single metabolic enzyme}
\begin{figure}[!ht]
\begin{center}
\includegraphics[width=8.3cm]{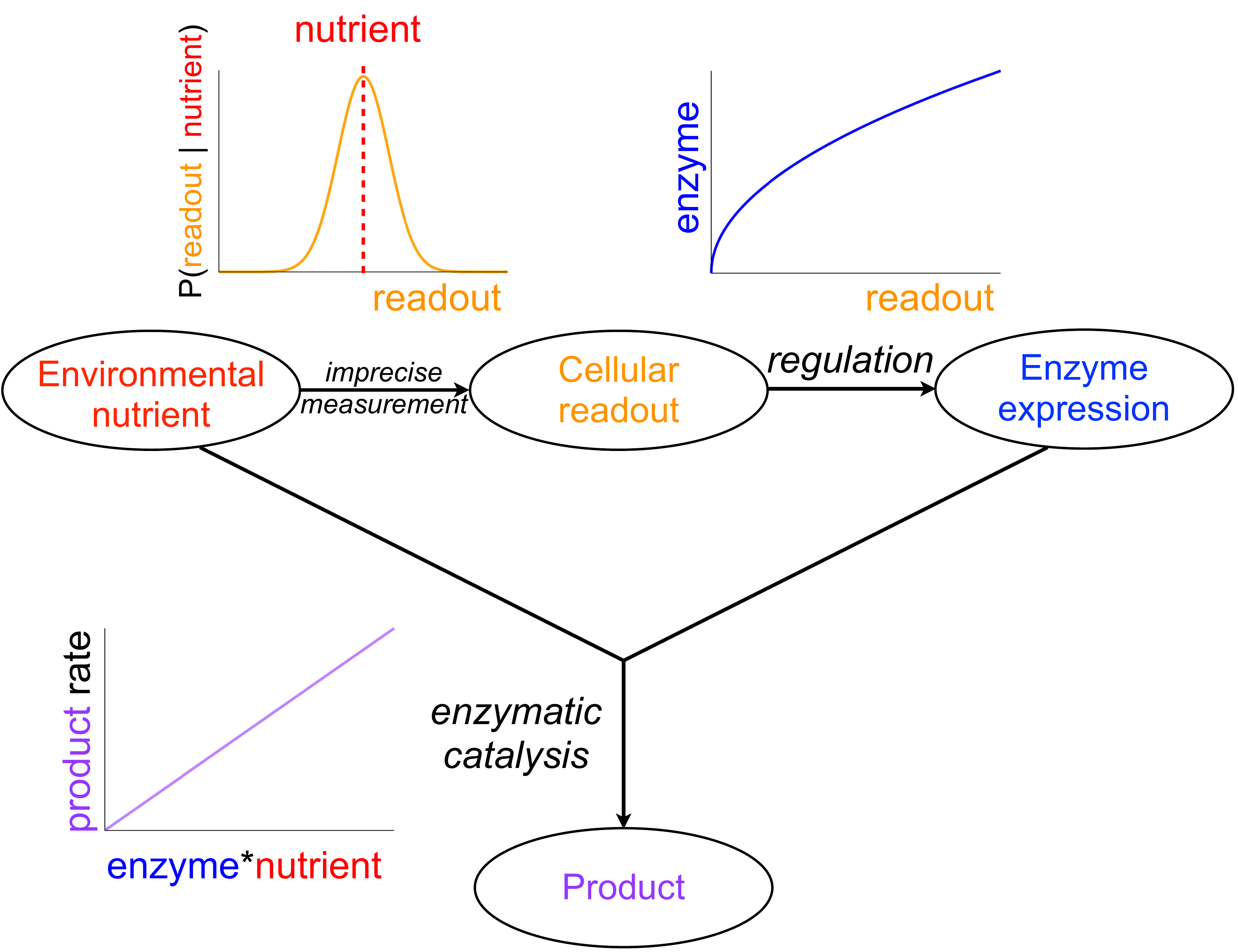}
\end{center}
\caption{{\bf Model system.} A time-varying environmental signal, the concentration of a nutrient, is read by the cell through a noisy process. Through regulation, the cell chooses an enzyme level, which interacts with the true nutrient concentration to produce product. In this work we focus on the optimization of the regulatory strategy, the choice of enzyme level as a function of the imperfect readout of nutrient concentration.}
\label{fig:Fig1}
\end{figure}

We consider the cellular task of responding to a time-varying stochastic environmental signal by regulating the expression of a single metabolic enzyme $E$ that metabolizes a nutrient $S$ directly into some useful downstream product $P$~\cite{Dekel:2005ey} (see Fig.~\ref{fig:Fig1}). We formulate the cell's task as implementing the regulatory strategy $e_{\rm opt}(s)$, a mapping of nutrient concentration $s$ to enzyme concentration $e$ that maximizes a payoff function $\payoff(e,s)$. $\payoff(e,s)$ quantifies the net payoff to the cell as the difference of a benefit $\benefit(e,s)$ and a cost $\cost(e)$. Initially we assume precise cellular measurement of the environment, namely the cell measures $s$ exactly.

The benefit $\benefit(e,s)$ reflects the downstream product generated by enzyme-catalyzed metabolism of the nutrient. Under Michaelis-Menten enzyme kinetics we propose a benefit function $\benefit(e,s) = \frac{e \, s}{K + s}$, for Michaelis constant $K$ and enzyme concentration $e$ in units of $V_{\rm max}$. When concentrations remain sufficiently low that the enzyme is in the unsaturated regime, the Michaelis-Menten benefit function becomes linear in both enzyme and nutrient levels, 
\begin{equation}
\benefit(e,s) = \f{e\, s}{K} \ . 
\label{equ:benefitUnsaturated}
\end{equation}

We model the enzyme production cost $\cost(e)$ as depending only on the current enzyme concentration $e$, reflecting the consumption of precursor molecules and energy in the synthesis of enzyme~\cite{Shachrai:2010}. In particular we adopt a simple cost function $\cost(e) = \costConst \, e^n$, $n>0$, a polynomial function of the current enzyme concentration $e$, where $n$ determines the convexity of the function.  (A strictly \emph{concave} function has $n<1$, whereas a strictly \emph{convex} function has $n>1$.) Different studies suggest that components of the lactose regulatory machinery may have convex~\cite{Dekel:2005ey} or concave~\cite{Eames:2012dm} costs across the expression range experimentally probed, and hence we explore how optimal regulatory strategies vary with cost convexity.

\subsection*{Precise measurement and the suitability of thresholding vs. graded response}
In this section we ask when should a cell threshold: when should it implement a discrete response or instead produce a graded response to environmental concentrations? We find that the relative convexity of the expression cost function produces a preference for either graded or switch-like regulatory strategies. 

\begin{figure}[!ht]
\begin{center}
\includegraphics[width=12.35cm]{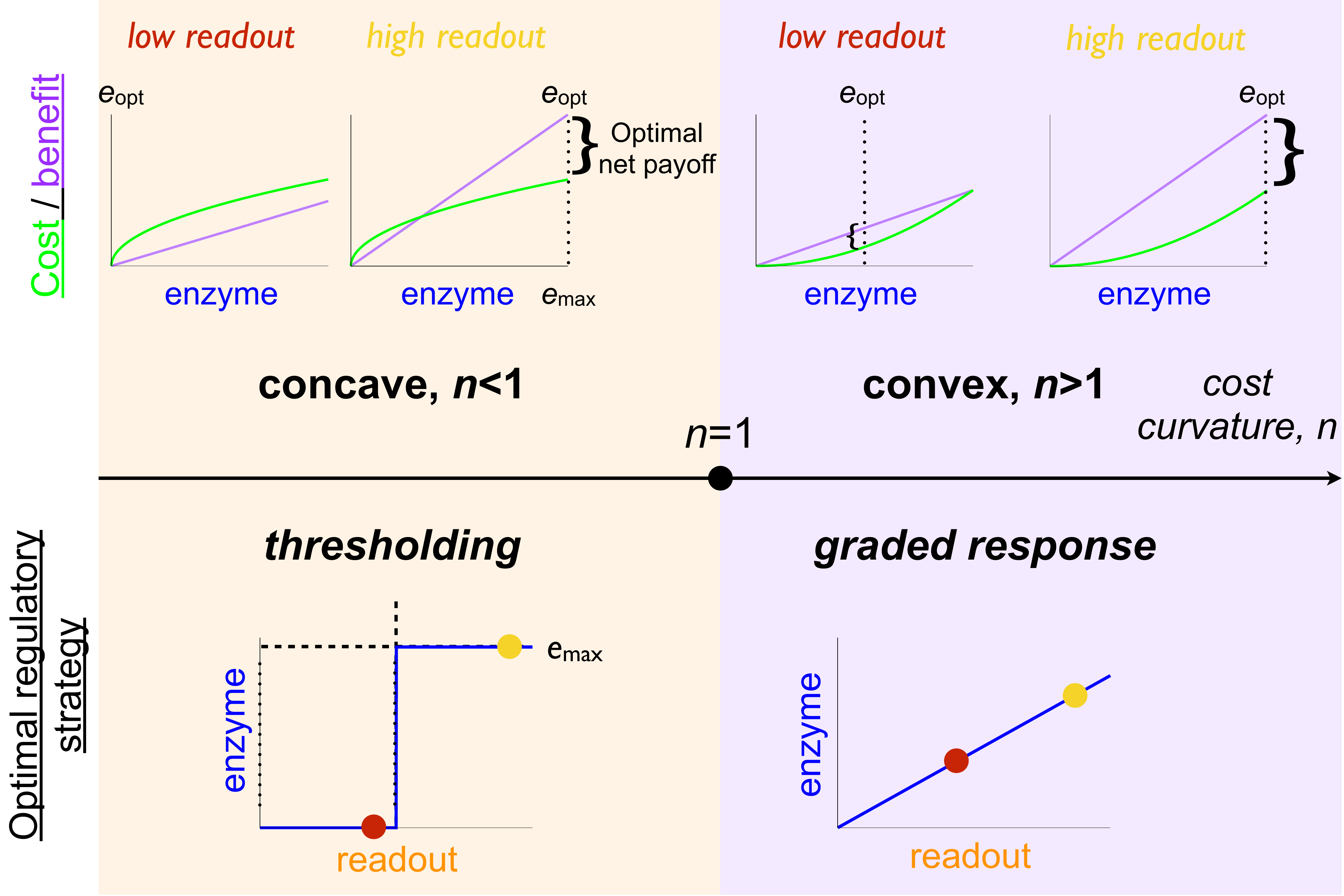}
\end{center}
\caption{{\bf Cost convexity relative to benefit produces preference for either thresholding or for graded response.} For a benefit function that is linear in nutrient concentration $s$ (purple curves in top panel) and a simple polynomial cost function $\costConst \, e^n$, concave cost ($n \le 1$, left column) implies an optimal enzyme expression level $\eOpt$ of either zero or the maximal enzyme level $\eMax$ (thresholding), whereas convex cost ($n>1$, right column) implies an optimal enzyme expression level that varies continuously with the cellular readout (graded response). Top row: costs (green curves) and benefits (purple curves) associated with an enzyme expression level for a given nutrient concentration. Bottom row: optimal regulatory strategy specifying a enzyme expression level for a given cellular readout.}
\label{fig:Fig2}
\end{figure}

For perfect sensing of the environment, the optimal regulatory strategy $\eOpt(s)$ is determined by maximizing the payoff function $\payoff(e,s)$ for each precisely-detectable nutrient level $s$. In the regime of strictly convex cost, $n>1$ (Fig.~\ref{fig:Fig2} right column), the optimal regulatory algorithm continuously tracks $s$ according to a graded response whose specific form is determined by the curvature of the cost function:
\begin{align}
\eOpt(s) = \left( \frac{s}{K\costConst \, n} \right)^{\frac{1}{n-1}}\ . 
\end{align}

For strictly concave enzymatic costs, $n < 1$ (Fig.~\ref{fig:Fig2} left column), the payoff function has no local maximum for non-negative $e$, and thus the optimal enzyme level must be on the boundary, either zero or $\eMax$ (the maximum level of enzyme that the cell can produce). For threshold nutrient concentration $\sThres \equiv K\costConst\, e^{n-1}_{\rm max}$, if $s > \sThres$, then the optimal regulatory strategy sets $\eOpt(s) = \eMax$, whereas when $s < \sThres$, the payoff function $\payoff(e,s)$ is negative for all $e$, so enzymatic production consumes more energy than it generates, and $\eOpt(s) = 0$. Thus the cell should switch between no enzyme production and maximal enzyme production whenever nutrient concentration $s$ crosses $\sThres$.

When nutrient concentration is relatively high, $s \not\ll K$, the benefit function adopts the Michaelis-Menten form. The benefit function becomes hyperbolic in $s$ but remains linear in $e$, so the solution again breaks into two qualitatively distinct scenarios of thresholding and graded response, depending on the convexity of the cost function (see Models). More generally, for any cost and benefit functions $\cost(e) \propto e^n$ and $\benefit(e) \propto e^m$ that are power laws of the enzyme concentration $e$, the optimal regulatory strategy will involve graded response whenever the cost function is strictly convex \emph{relative to} the benefit function, $n > m$, and thresholding whenever cost is strictly concave relative to benefit, $n < m$ (see Models).

In this way, optimal regulatory algorithms with perfect measurement fall into two qualitative classes: for a cost function strictly convex relative to benefits, the cell should track the environment with a graded regulatory strategy; and for a cost function strictly concave relative to benefits, the cell should perform thresholded switching between on and off enzyme states. Thus, a discrete or continuous regulatory strategy is optimal depending on the relative curvatures of the enzymatic cost and benefit functions.

\subsection*{Imperfect measurement and the value of a Bayesian response strategy}
In this section we ask when there is a fitness advantage to implementing sophisticated Bayesian decision rules, which combine information from present measurement and prior knowledge of environmental statistics. We find such an advantage in contexts of medium measurement imprecision relative to environmental variability, when uncertainty is sufficiently low that individual measurements have informational value, but sufficiently high that prior knowledge is also useful.  

Cells measure the concentration of environmental nutrients through protein sensors (often membrane-bound receptors). These sensors exist in small copy numbers and are subject to strong thermal conformational fluctuations, thus the cellular measurement apparatus operates stochastically rather than deterministically, providing imperfect measurements of nutrient concentrations~\cite{Berg:1977bp,Libby:2007fw}. In this way, instead of responding to $s$, the true concentration of an environmental nutrient, the cell responds to $s^*$, a corrupted measurement or readout of $s$.  We now ask how a cell can optimally regulate enzyme level based upon imperfect knowledge of the environment. 

The cell's regulatory strategy must depend only upon measured concentration $s^*$, but the cell's payoff $\payoff(e,s)$ will depend upon the true concentration of nutrients. The nutrient sensor is characterized by the conditional measurement distribution, $P(s^*|s)$, the probability of the sensor measuring a nutrient level $s^*$ given a true nutrient concentration $s$. The optimal regulatory strategy $\eOpt(s^*)$ maximizes the expected payoff function $\payoff(e,s^*) \equiv E[ \payoff(e,s) | s^*]$ given a measurement $s^*$, averaging over the different possible true nutrient concentrations $s$. Note that in this optimization we assume that fitness only depends on cost and benefit \emph{averages}, not on their variances or higher-order moments. We initially consider environments that vary but are uncorrelated in time, and introduce the prior environmental distribution $P(s)$, the probability of the nutrient concentration at any instant in time. In this section we explore the optimal regulatory strategy for specific forms of the payoff function, environmental prior of nutrient concentrations, and conditional measurement distribution.

For the unsaturated enzyme benefit function [Eq.~\eqref{equ:benefitUnsaturated}] with strictly convex costs, $n>1$, the optimal enzyme level for a given measured nutrient concentration $s^*$ is:
\begin{equation}
\eOpt(s^*) = \left(\frac{E[ s | s^* ]}{K\costConst \, n}\right)^{\frac{1}{n-1}} \ .
\end{equation}
Due to the linear dependence of this benefit function on nutrient concentration, the optimal response now depends upon $E[ s | s^* ]$, the expectation of the environmental nutrient concentration $s$ given a measurement $s^*$. Via Bayes' rule this expectation depends upon both the prior distribution of nutrient concentrations $P(s)$ and the conditional measurement distribution $P(s^*|s)$:
\begin{equation}
E[ s | s^* ] = \int \md s \, \f{P(s^*|s) P(s)}{P(s^*)} \, s \ .
\end{equation}
In the presence of measurement noise, Bayes' rule motivates consideration of environmental statistics, encoded in $P(s)$, in the maximization of $\payoff(e,s^*)$, through calculation of the cell's expectation $E[s|s^*]$ of $s$ given a measured $s^*$. The prior distribution, $P(s)$, is presumably learned over evolutionary timescales. Several previous studies have postulated a role for Bayesian inference in nutrient sensing~\cite{Libby:2007fw,Kobayashi:2010io}; in our framework, Bayesian inference emerges as a result of maximization of expected enzymatic benefit averaged over realizations of a stochastic environment.

Expectations preserve convexity, so the basic results under perfect measurement are preserved: \emph{e.g.}, in the strictly concave cost regime where $n < 1$, a switch-like response is again optimal, now depending on the expected nutrient level given the measurement. Henceforth we assume strictly convex costs, $n > 1$, and an unbiased Gaussian measurement error, and we examine optimal enzymatic regulatory strategies for different environments specified by the nutrient distribution $P(s)$.

\subsubsection*{Unimodal nutrient distribution}
First we assume a simple Gaussian distribution of nutrient concentrations. Straightforward calculation reveals that for mean nutrient level $\avUni$,
\begin{equation}
E[ s | s^* ] = \f{1}{1+r} s^* + \f{r}{1 + r} \avUni \ ,
\label{equ:eUnimodal}
\end{equation}
where $r$ is the dimensionless ratio of variances of conditional nutrient distributions and measurement errors:
\begin{equation}
r \equiv \f{\sM^2}{\sS^2} \ . 
\end{equation}
In this context $r$ is the inverse of the signal-to-noise ratio. The optimal enzyme level is graded with respect to the measurement $s^*$:
\begin{equation}
\eOpt(s^*) = \left[ \f{\f{1}{1+r} s^* + \f{r}{1 + r} \avUni }{K \costConst \, n} \right]^{\f{1}{n-1}} \ .
\end{equation}

\begin{figure}[!ht]
\begin{center}
\includegraphics[width=12.35cm]{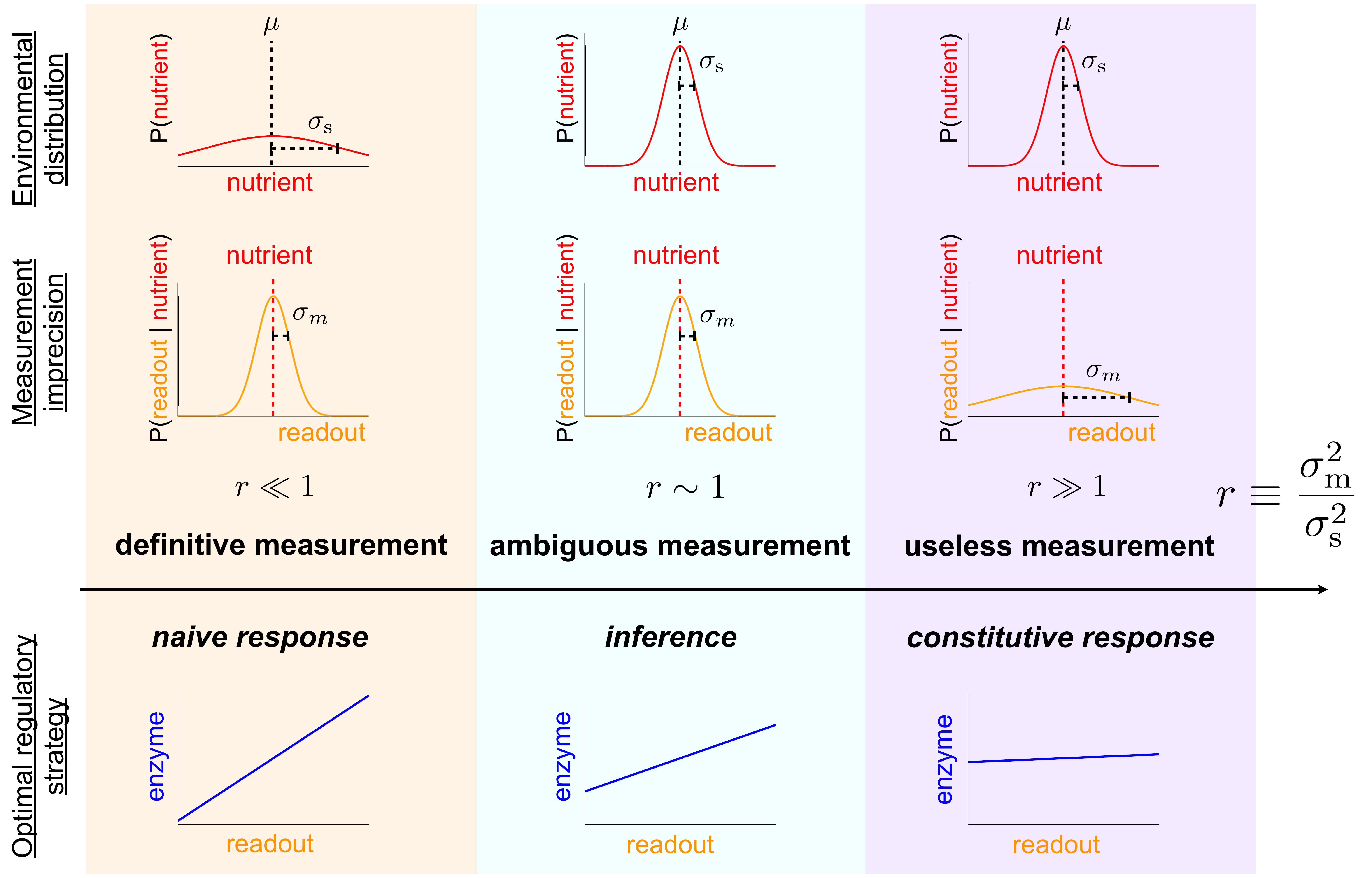}
\end{center}
\caption{{\bf Increasing measurement noise shifts the optimal strategy from naive response to constitutive response.}  For a quadratic cost function ($n=2$) and relatively slow environmental dynamics, the dimensionless ratio $r \equiv \sM^2/\sS^2$ of the measurement imprecision $\sM^2$ (middle row) and the environmental variation $\sS^2$ (top row) determines the preference for different regulatory strategies [see Eq.~\eqref{equ:eUnimodal}]. Low relative measurement noise ($r \ll 1$, left column) leads to a preference for naive response; high relative measurement noise ($r \gg 1$, right column) produces a preference for constitutive response; and the intermediate case ($r \sim 1$, middle column) leads to a preference for more sophisticated inference incorporating both prior knowledge and the current measurement of the environment. Top row: distribution of possible environmental nutrient concentrations around the mean $\mu$. Middle row: distribution of cellular readouts given a particular nutrient concentration (red dotted line).}\label{fig:Fig3}
\end{figure}

When measurement uncertainty is small compared to environmental variability, $\sM^2 \ll \sS^2$ and hence $r \ll 1$ (``definitive measurement,'' Fig.~\ref{fig:Fig3} left column), the cell can confidently distinguish between many different common nutrient concentrations on the basis of a single measurement, with the environmental prior providing negligible additional information. The expected nutrient level is the measurement, $E[ s | s^* ] \approx s^*$, and hence the optimal strategy involves naive response to the measurement. Conversely, for high relative measurement uncertainty, $r \gg 1$ (``useless measurement,'' Fig.~\ref{fig:Fig3} right column), measurement provides negligible information not already contained in the environmental prior distribution. The expectation is the mean of the prior, $E[ s | s^* ] \approx \avUni$, corresponding to an optimal strategy of constitutive expression, \emph{i.e.,} unresponsiveness to changing measurements. In the intermediate regime, $r \sim 1$ (``ambiguous measurement,'' Fig.~\ref{fig:Fig3} middle column), the measurement provides some useful information but is not dispositive, so one updates the prior mean by the measurement, with relative weightings depending on the relative variances of nutrient concentrations $\sS^2$ (Fig.~\ref{fig:Fig3} top row) and measurement errors $\sM^2$ (Fig.~\ref{fig:Fig3} middle row). This produces an optimal strategy of a \emph{non-degenerate} Bayesian decision rule, one that makes use of both prior information and the current measurement. Notice that the quantitative level of optimal enzyme expression is determined by statistical properties of the environment: for $r \gg 1$, the optimal expression level is set by the mean of the environmental nutrient concentration. 

\subsubsection*{Bimodal nutrient distribution}
\begin{figure}[!ht]
\begin{center}
\includegraphics[width=12.35cm]{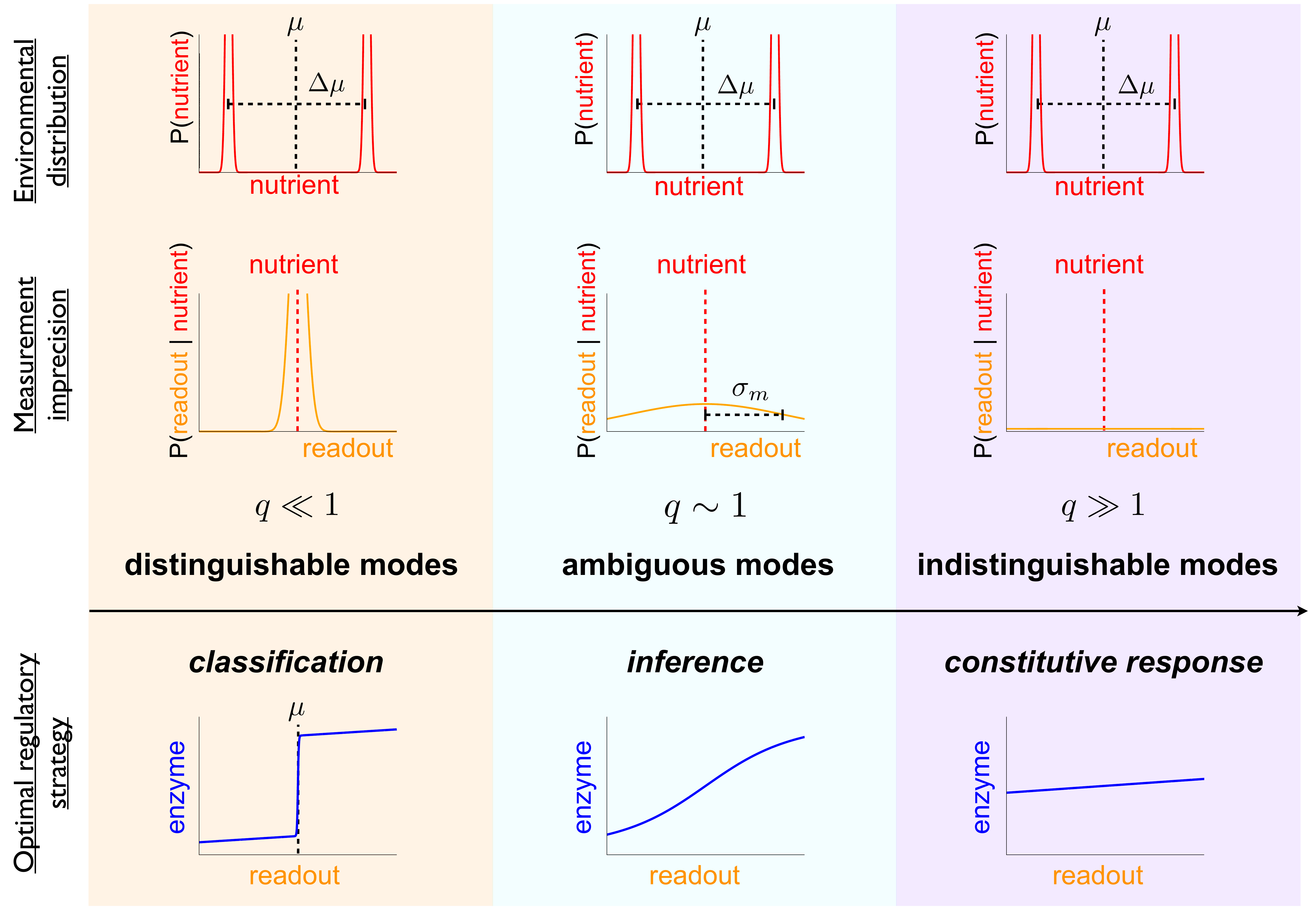}
\end{center}
\caption{{\bf In a bimodal environment, increasing measurement noise shifts the optimal strategy from classification to constitutive response.} For a quadratic cost function, tight distribution within each environmental mode (such that $r \gg 1$), and relatively slow environmental dynamics between distinct environmental modes (with mode separation $\Delta \mu$), the dimensionless ratio $q \equiv  \sM^2 / (\Delta\mu[\half\Delta\mu + \sS])$ determines the preference among regulatory strategies [see Eq.~\eqref{equ:bimodalStrategy}]. High relative measurement noise ($q \gg 1$, right column) leads to a preference for constitutive response; low relative measurement noise ($q \ll 1$, left column) produces a preference for classifying the environment into the most likely among the two modes; and the intermediate case ($q \sim 1$, middle column) produces a preference for non-degenerate Bayesian inference.}
\label{fig:Fig4}
\end{figure}

We now examine an environmental nutrient distribution with more complex structure, specifically an environment that fluctuates between two dominant conditions, one of abundant nutrient and one of scarce nutrient (Fig.~\ref{fig:Fig4}). Concretely, we assume $P(s)$ is an equiprobable mixture of two Gaussians, each with the same variance $\sS^2$, with means separated by $\Delta\mu$ (Fig.~\ref{fig:Fig4} top row), and overall environmental mean $\avBin$. Integration shows that the posterior mean of the true environmental concentration $s$, conditioned on the measurement $s^*$, is
\begin{align}
E[ s | s^* ] = \f{1}{1+r} s^* + \f{r}{1+r}\left[ \avBin + \tfrac{1}{2}\Delta\mu \tanh \f{\Delta\mu(s^*-\avBin)}{2(\sM^2 + \sS^2)} \right] \ . 
\label{equ:bimodalStrategy}
\end{align}
When measurement uncertainty is small compared to environmental variability within a given mode, $r \ll 1$, the expectation is the measurement, $E[ s | s^* ] \approx s^*$. 

Where measurement uncertainty is large compared to environmental variability within a given mode, $r \gg 1$, the cell can only hope to distinguish between modes, not specific nutrient levels within a mode. In this context we highlight three qualitatively distinct regimes (Fig.~\ref{fig:Fig4}) demarcated by the dimensionless parameter $q \equiv \sM^2 / (\Delta\mu[\half\Delta\mu + \sS])$. $q$ is the ratio of the measurement uncertainty to the product of the separation $\Delta\mu$ between the two mean nutrient levels and the typical distance $s^*-\Delta\mu \sim \half\Delta\mu+\sS$ of a measurement to the mean. 

Larger $q$ corresponds to a wider range of measurements that leave some ambiguity about which mode the environment is in: when $q \gg 1$ (``indistinguishable modes,'' Fig.~\ref{fig:Fig4} right column), measurement is insufficiently precise to distinguish between the two modes, and hence the optimal strategy produces constitutive enzyme expression at a level corresponding to the mean value $\avBin$ of the environment. In the opposite limit, $q \ll 1$ (``distinguishable modes,'' left column), measurement is relatively precise compared to the separation between the modes, and hence essentially all possible measured nutrient levels strongly implicate one or the other mode. Thus the optimal strategy is classification, choosing either of the mean nutrient concentrations $\mu_{\rm L}$ or $\mu_{\rm R}$,
\begin{align}
E[ s | s^* ] = \Bigg\{
        \begin{array}{ll}
            \mu_{\rm L}, & \quad s^* < \avBin \vspace{0.0ex}\\
            \mu_{\rm H}, & \quad s^* > \avBin
        \end{array}
    \ .
\end{align}
In the intermediate regime, $q \sim 1$ (``ambiguous modes,'' middle column), the modes are moderately distinguishable but most measurements are not strongly indicative of one mode or the other, so the optimal strategy calls for more nuanced inference. Fig.~S1 depicts optimal regulatory strategies across varying $r$ and $q$. These optimal strategies can also be generalized to a \emph{multimodal} Gaussian mixture model (see Models). 

In this way, a stochastic environment imposes structure on the optimal sensing strategy through estimation of nutrient levels based on environmental statistics. Prior knowledge of the multimodal nature of the environmental nutrient distribution (\emph{e.g.}, producing only either scarcity or abundance) leads to an optimal regulatory strategy that infers the environmental state from a measured concentration of nutrient. When measurement noise is low estimation is not required, and when measurement noise is very high estimation is not possible; in the intermediate regime, optimal regulatory strategies are non-degenerate Bayesian decision rules. 

In addition to specifying the broad structure of the optimal sensing strategy in a bimodal environment, Eq.~\eqref{equ:bimodalStrategy} relates the quantitative architecture, and hence underlying biochemical parameters, of the optimal sensing apparatus to statistical properties of the environment. For example, the optimal sensing strategy is to threshold the readout into a discrete on or off response in the regime $r \gg 1$ and $q \ll 1$. Quantitatively, the mean level $\avBin$ of the nutrient $s$ across environmental realizations sets the optimal location of the switch threshold. Additionally, for $r \gg 1$ and varying $q$ the optimal strategy adopts the sigmoidal shape of the $\tanh$ function where the steepness or cooperativity of the optimal thresholded response is determined by the ratio of the separation between the two environments $\Delta\mu$ and the summed environmental and measurement variances $(\sM^2+\sS^2)$. The thresholding strategy could be implemented using sigmoidal responses (commonly arising in biochemical networks), where the statistical properties of the environment and measurement apparatus set the biochemical parameters, including dissociation constant and Hill coefficient, that optimize the thresholding properties of the switch~\cite{Gunawardena:2005jm, Huang:1996vm}. In this way, the model suggests a fitness benefit for internalizing environmental structure in the value of specific biochemical parameters, in agreement with recent theoretical work analyzing the fundamental connections between energetic efficiency and predictive efficiency~\cite{Still:2012ta}. 

Ref.~\cite{Libby:2007fw} analyzed Bayesian decision rules in an environment that is a mixture of two sharply-peaked Gaussians in log space, representing high nutrient and low nutrient concentrations, respectively. By continuously parametrizing both the statistics of the environment as well as measurement imprecision, our framework generalizes these results to environments that switch more gradually. Like \cite{Libby:2007fw}, we find that the optimal sensing strategy is a switch-like strategy when the environment has a sharp two-state structure. Additionally, our generalized framework allows continuous analysis of optimal regulatory strategy while titrating the environmental structure from one that is sharply peaked to one with more continuous variation.

\subsection*{Dynamic environments and the value of memory}
In this section we ask when should a cell remember: when does a cell benefit from retaining memory of past environmental states? In dynamic contexts, we find that retaining memory produces a fitness advantage for intermediate levels of measurement imprecision, where measurement is sufficiently precise to constrain possible environmental states, but still noisy enough that inference benefits from combining present and past measurements.  

So far, we have implicitly assumed that a cell does not retain any memory of specific past measurements. But an environment with temporal correlations that persist longer than cellular measurement intervals will reward more sophisticated inference algorithms. Here we address how a cell can optimally combine sequential measurements of a nutrient signal in time to regulate the level of the corresponding metabolic enzyme. 

In particular, we seek a regulatory strategy $\eOpt(s_{\nT}^*,s_{\nT-1}^*)$ that maximizes the value of the payoff function $\payoff(e,s_{\nT})$ averaged over possible current nutrient concentrations $s_{\nT}$, where now the regulatory strategy depends in principle on both current ($s_{\nT}^*$) and past ($s_{\nT-1}^*$) measurements of the nutrient signal. We find qualitatively similar features to the simpler uncorrelated case, namely the effect of relative cost convexity on the preference for graded or switch-like responses, and the transitioning between naive response, Bayesian response, and constitutive response on the basis of the ratio of relevant noises. However in this dynamic context, the intermediate case of a non-degenerate Bayesian decision rule depends on past measurements.  

We assume that the environmental dynamics are Markovian, and that successive measurements depend only on the current true nutrient  via a time-invariant measurement distribution $P(s_{\nT}^*|s_{\nT})$. For the specific payoff function $\payoff(e_{\nT},s_{\nT}) = \f{e_{\nT}}{K} \, s_{\nT} - \costConst\, e_{\nT}^n$, the expected payoff is
\be
\payoff(e_{\nT},s_{\nT}^*,s_{\nT-1}^*) = \f{e_{\nT}}{K}\, E[s_{\nT}|s_{\nT}^*, s_{\nT-1}^*] - \costConst\, e_{\nT}^n \ .
\ee

For further concreteness, we specify a mean-reverting diffusive environment with conditional nutrient distribution $P(s_{\nT}|s_{\nT-1}) = f(s_{\nT}; \avDyn + a [s_{\nT-1} - \avDyn], [1-a^2]\sS^2)$, where $a$ ($0 \le a \le 1$) is the environmental persistence and $f(x; m,\sigma^2)$ is a normal distribution for $x$ with mean $m$ and variance $\sigma^2$. Such an environment executes a random walk in nutrient concentration space with constant marginal distribution $P(s_{\nT}) = f(s_{\nT}; \avDyn, \sS^2)$ and correlation time $-1/\ln a$. Hence the smaller $a$ is, the quicker the nutrient concentration reverts to its mean and hence the more rapidly correlation decays between nutrient concentrations at different time points. With the same Gaussian measurement error as before, straightforward integration leads us to an expected nutrient concentration, given the current and previous measurements,
\begin{align}
E[s_{\nT}|s_{\nT}^*,s_{\nT-1}^*] = \frac{[(1-a^2) + r]s_{\nT}^* + a\,r \, s_{\nT-1}^* + [(1-a)r + r^2] \avDyn }{(1-a^2) + 2 r + r^2} \ .
\label{equ:dynamicStrategy}
\end{align}
The linear mean-reversion, quadratic diffusion, and quadratic measurement errors ensure that this estimate is precisely that of a Kalman filter~\cite{Kalman:1960ii,WelchBishop}.

\begin{figure}[!ht]
\begin{center}
\includegraphics[width=12.35cm]{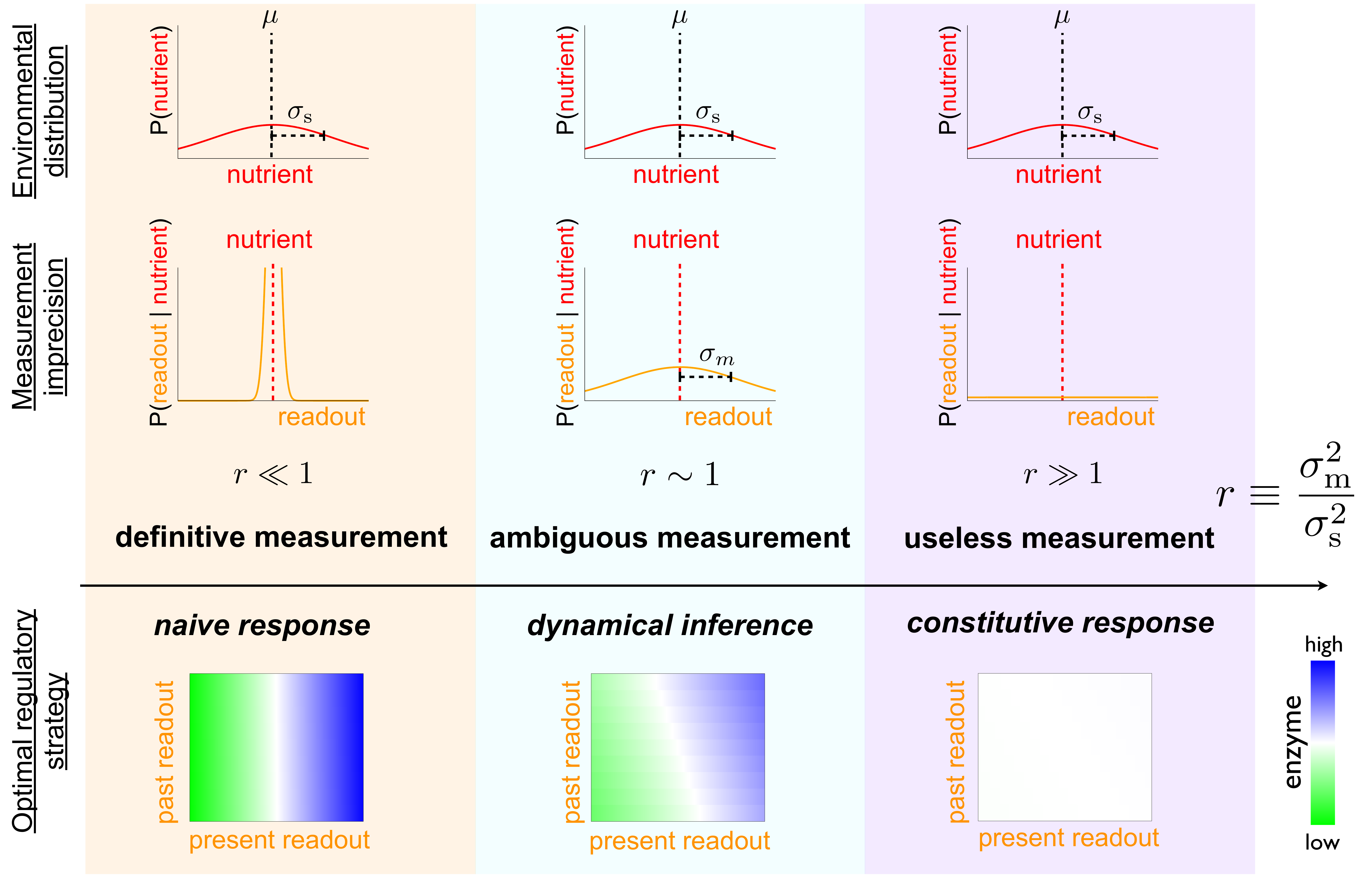}
\end{center}
\caption{{\bf In a rapidly changing environment, the value of memory peaks at intermediate measurement noise.} For a quadratic cost function and environmental changes on timescales comparable to cellular response, the dimensionless ratio $r$ determines the preference among different regulatory strategies [see Eq.~\eqref{equ:dynamicStrategy}]. High relative measurement noise ($r \gg 1$, right column) leads to a preference for constitutive response; low relative measurement noise ($r \ll 1$, left column) produces a preference for naive response to the present measurements; and the intermediate case ($r \sim 1$, middle column) produces a preference for dynamic Bayesian inference that takes into account both present and past measurements. In the heat maps (bottom row), blue represents high levels of enzyme and green represents low.}
\label{fig:Fig5}
\end{figure}

When the conditional variance of nutrients dwarfs the measurement error, $r \ll 1$ (Fig.~\ref{fig:Fig5} left column), the best inference is the current measurement $s_{\nT}^*$; when measurement imprecision is relatively high, $r \gg 1$ (right column), the best inference is the nutrient mean $\avDyn$; and in the intermediate regime, $r \sim 1$ (middle column), a dynamic Bayesian decision rule combines the two along with information from the previous measurement $s_{\nT-1}^*$. Fig.~S2 depicts optimal regulatory strategies for varying levels of $r$ and environmental persistence $a$.

Cellular memory of a past measurement $s_{\nT-1}^*$ can be instantiated in forms such as epigenetic chromatin modification~\cite{Ginsburg:2009ul}, long-lived proteins~\cite{Sigal:2006bd}, and even particular network topologies~\cite{Lisman:1985tp}, and indeed such a dynamic Bayesian decision rule as described above can be implemented by noisy receptors and intracellular kinetics featuring dual positive feedback~\cite{Kobayashi:2010io}. Inference of the current nutrient concentration can benefit from incorporation of information from even earlier measurements ($s_{\nT-2}^*,s_{\nT-3}^*,\ldots$), and the above derivation generalizes trivially, but the resulting expressions rapidly grow cumbersome (see Models). In multicellular contexts with environmental dynamics relatively rapid compared to regulatory timescales, stochastic enzymatic expression can provide additional fitness advantages~\cite{Thattai:2004tk,Kussell:2005dg}.

Eq.~\eqref{equ:dynamicStrategy} suggests that optimal regulatory strategies internalize the temporal structure of the environment in the signal-processing apparatus. Namely, $a$ is related to the correlation time of the environment (see Models), and $r$ depends upon the environmental variance, so that an optimal regulatory strategy requires learning through evolution the correlation structure of the environment, the feasibility of which has been demonstrated by recent microevolution studies~\cite{Tagkopoulos:2008ct,Mitchell:2009el}.

\subsection*{Conclusions}
In the analysis presented here, measurement noise and environmental structure interact to determine the optimal regulatory strategy. In this work we specifically find that: (i) convexity of enzyme expression cost, relative to benefit, influences preferences for thresholding or graded responses; (ii) intermediate levels of uncertainty call for a sophisticated Bayesian decision rule that combines prior information with new measurement; and (iii) in dynamic contexts, intermediate levels of uncertainty call for retaining memory of the past. 

The perspective adopted here provides a decision-theoretic framework for interpreting existing biomolecular signal processing algorithms, by relating optimal response to environmental and cellular statistics in a novel yet intuitive manner. It is easily extensible to provide computational tools for predicting optimal regulatory strategies in complex environments where correlations are derived directly from ecological data. 
The framework represents a natural classification system that, through continuous variation of dimensionless parameters, relates a range of regulatory strategies that at first glance appear qualitatively distinct. Further exploration of parameter space (for example, see Fig.~S1) may suggest novel forms distinct from commonly-studied regulatory strategies such as thresholding.

Our work motivates new experiments that compare the fitness of signal-processing strategies in different regimes of environmental structure and sensing noise. For example, we predict that in a bimodal environment, varying between starvation and nutrient-rich conditions, when measurements are very imprecise (because of low copy number receptors) a cell constitutively expressing the corresponding metabolic enzyme will outperform a cell regulating enzyme expression. Experiments to test these ideas could compare, in rapidly-changing microfluidics environments, the fitness of synthetic nutrient response pathways designed to implement either constitutive or graded response, with measurement noise titrated via differing steady-state receptor copy numbers due to high- or low-copy number plasmids.

\section*{Models}
Our model system is an enzyme $E$ that metabolizes a nutrient $S$ into some useful downstream product $P$, according to the reaction scheme
\begin{align}
{\rm E} + {\rm S} \rightarrow {\rm E} + {\rm P} \ .
\end{align}
We formulate the cell's regulatory task as choosing the concentration of enzyme that maximizes a function $\payoff(e,s)$ quantifying the net payoff to the cell given both the enzyme concentration $e$ and the environmental nutrient concentration $s$:
\begin{align}
\payoff(e,s) &= \benefit(e,s) - \cost(e) \ .
\end{align}
A \emph{regulatory strategy} $\eOpt(s)$ specifies the enzyme level that maximizes the net payoff $\payoff(e,s)$. 

Under Michaelis-Menten enzyme kinetics we propose a benefit function $\benefit_{\rm MM}(e,s) = e \, s / (K + s)$, for Michaelis constant $K$ and enzyme concentration $e$ in units of $V_{\rm max}$. In the limit of small nutrient concentration and hence unsaturated enzyme kinetics, $s \ll K$, this benefit function simplifies to a linear function of $e$ and $s$, $\benefit_{\rm unsat}(e,s) = e \, s / K$. We adopt a simple cost function $\cost(e) = \costConst \, e^n$. 

We initially consider a model where the environment is changing in an uncorrelated fashion so that at any instant in time, the cell is exposed to the nutrient at concentration $s$ with probability $P(s)$. The cell does not have direct access to $s$, but rather it measures through noisy protein sensors an estimated nutrient concentration $s^*$. The aim of our framework is to derive an expression for the optimal expression level $\eOpt$ given a measured $s^*$, a function $\eOpt(s^*)$ that maximizes the average value of the payoff function $\payoff(e,s)$. (We assume that fitness does not depend on the payoff variance or higher-order moments.) For simplicity, we assume that the cell can respond to the measured nutrient concentration faster than the typical timescales for environmental change. 

First, we find the average value of the payoff function conditioned on $s^*$ by deriving an expected payoff function $\payoff(e,s^*)$ given a measured $s^*$, averaging over the possible nutrient concentrations $s$: 
\begin{subequations} 
\begin{align}
\payoff(e,s^*) &\equiv E[ \payoff(e,s) | s^*] \\
&= \int \md s \, \payoff(e,s) P(s|s^*) \\
&= \int \md s \, \payoff(e,s) \frac{P(s^*|s) P(s)}{P(s^*)} \\
&= \int \md s \, \payoff(e,s) \frac{P(s^*|s) P(s)}{\int \md s' P(s^*|s') P(s')} \ .
\end{align} 
\end{subequations} 
This expected payoff depends upon the environmental statistics, $P(s)$, as well as the conditional distribution, $P(s^*|s)$, of measuring $s^*$ given the actual concentration $s$. The third line follows from Bayes' rule, and the fourth line follows from the law of total probability,
\begin{equation}
P(s^*) = \int \md s \, P(s^*|s) P(s) \ .
\end{equation}

Maximizing $\payoff(e,s^*)$ with respect to $e$ produces an expression for $\eOpt$, the optimal level of enzyme expression $e$, for each measurement $s^*$: 
\begin{equation}
\eOpt(s^*) \equiv \text{argmax}_{e} \ \payoff(e,s^*) \ . 
\end{equation}
We call this function $\eOpt(s^*)$ the optimal \emph{regulatory strategy}. 

For our specified payoff function with unsaturated enzyme kinetics, 
\begin{subequations} 
\begin{align}
\payoff(e,s) &= \f{e}{K} \, s- \costConst \, e^n\\
\payoff(e,s^*) &= \int \md s \, \l( \f{e}{K} \, s - \costConst \, e^n\r) P(s|s^*) \\
&= \f{e}{K} \, E[ s | s^* ] - \costConst \, e^n \ .
\end{align} 
\end{subequations} 

In the name of simplicity, tractability, and interpretability, this model contains a number of simplifying assumptions: the cell can sense and respond to a signal on timescales faster than those on which the environment varies; the metabolic benefit is linear in the enzyme concentration; system cost is only a function of the current level of enzyme; all regulatory mechanisms are equally costly, regardless of their steady-state energy requirements, number of required components, or overall complexity; the cell can set a deterministic enzyme level in response to a given readout level; and we only consider a single enzyme and single nutrient. We also assume simple functional forms throughout this framework in order to derive analytic results, though the qualitative character of these results should be robust to modest variation of the model details.

\subsection*{Precise measurement and the suitability of thresholding vs. graded response}
We start with the case of perfect detection, where we immediately see that $E[ s | s^* ] = s^*$, and hence in the strictly convex cost regime, $n>1$, the optimal enzyme level is
\begin{equation}
\eOpt(s) = \left(\frac{s}{K\costConst \, n}\right)^{\frac{1}{n-1}} \ .
\end{equation}
By contrast, in the strictly concave cost regime, $n < 1$,
\begin{align}
\eOpt(s) = \bigg\{
        \begin{array}{ll}
            \eMax, & \quad s > \f{K\costConst}{e^{1-n}_{\rm max}} \\
            0, &\quad  s < \f{K\costConst}{e^{1-n}_{\rm max}} 
        \end{array}
    \ .
\end{align}

\subsubsection*{Michaelis-Menten kinetics}
For the full Michaelis-Menten benefit model, the benefit remains linear in $e$, so the solution again breaks into two qualitatively distinct scenarios of thresholding and graded response. 
For $n>1$,
\begin{align}
\eOpt(s) = \left[ \frac{s}{\costConst\, n(K+s)} \right]^{\frac{1}{n-1}} \ . 
\end{align}
Again, when $n < 1$, the payoff function is always an increasing function of enzyme level, so that 
\begin{align}
\eOpt(s) = \Bigg\{
        \begin{array}{ll}
            \eMax, & \quad s > \frac{K}{\f{e^{1-n}_{\rm max}}{c} - 1} \vspace{0.0ex}\\
            0, &\quad  s \leq \frac{K}{\f{e^{1-n}_{\rm max}}{c} - 1}
        \end{array}
    \ .
\end{align}

\subsubsection*{More general benefit function}
More generally, for any cost and benefit functions that are power laws of the enzyme concentration $e$, the payoff function will be
\be
F(e,s) = b \, s \, e^m - c \, e^n \ , 
\ee
with $n>0$ and $m>0$ reflecting increasing costs and benefits, respectively, with increasing enzyme level. For $n \ne m$ the payoff function has zero slope at
\be
e = \l( \f{b \, m \, s}{c \, n} \r)^{\f{1}{n-m}} \ .
\ee
If also $n \ne 1$ and $m \ne 1$, then the second derivative at the unique nonzero local optimum is
\be
\f{\pa^2 F(e,s)}{\pa e^2} \Big|_{e_{\rm opt}} = \l[ \f{(b \, m \, s)^{n-2}}{(c \, n)^{m-2}} \r]^{\f{1}{n-m}} \l( m - n \r) \ , 
\ee
which is positive for $n < m$ and negative for $n > m$. Thus the optimal regulatory strategy will involve graded response whenever the cost function is strictly convex \emph{relative to} the benefit function, $n > m$, and thresholding whenever cost is strictly concave relative to benefit, $n < m$.

\subsection*{Imperfect measurement and the value of Bayesian response strategies}
Henceforth, instead of perfect detection we assume an unbiased Gaussian error, whereby $s^*$ is Gaussian-distributed with mean equal to the true concentration of the nutrient $s$ and variance $\sM^2$,
\begin{equation}
P(s^*|s) = f(s^*; s, \sM^2) \ ,
\end{equation}
where $f(x; m,\sigma^2)$ is a normal distribution for $x$ with mean $m$ and variance $\sigma^2$. 

Local optima are found by differentiating with respect to $e$: 
\begin{equation}
0 = \frac{\md E[\payoff(e,s) | s^* ]}{\md e} \bigg|_{e=\eMax} =  \f{E[ s | s^* ]}{K} - n \, \costConst \, \eMax^{n-1} \ ,
\end{equation}
giving for strictly convex costs, $n>1$: 
\begin{equation}
\eOpt(s^*) = \left(\frac{E[ s | s^* ]}{K\costConst \, n}\right)^{\frac{1}{n-1}} \ .
\end{equation}
We are optimizing the expected payoff, without any concern for variance or higher-order moments of the payoff, which means that the optimal response in a stochastic environment is the same as the optimal response in the deterministic case, but $s^*$ is replaced by $E[ s | s^* ]$. 

For our specified payoff function, in the strictly convex cost regime, $n>1$, the optimal enzyme level for a given measured nutrient concentration $s^*$ is:
\begin{equation}
\eOpt(s^*) = \left(\frac{E[ s | s^* ]}{K\costConst \, n}\right)^{\frac{1}{n-1}} \ .
\end{equation}
Due to Bayes' rule this expectation $E[ s | s^* ]$ depends upon both the conditional measurement distribution $P(s^*|s)$ and the environmental structure $P(s)$:
\begin{equation}
E[ s | s^* ] = \int \md s \, \f{P(s^*|s) P(s)}{P(s^*)} \, s \ .
\end{equation}
In the strictly concave cost regime, $n < 1$, a switch-like response is again optimal:
\begin{align}
\eOpt(s^*) = \Bigg\{
        \begin{array}{ll}
            \eMax, & \quad E[s|s^*] > K \costConst\, e^{n-1}_{\rm max} \vspace{0.0ex}\\
            0, &\quad  E[s|s^*] \leq K \costConst\, e^{n-1}_{\rm max} 
        \end{array}
    \ .
\end{align}

\subsubsection*{Uniform nutrient distribution}
A uniform probability of nutrient levels corresponds to an uninformative prior, essentially a constant $P(s)$. Given the lack of any prior information about $s$, $E[ s | s^* ] = s^*$ and thus the optimal enzyme level is unchanged from the case of perfect detection.  

\subsubsection*{Unimodal nutrient distribution}
Here we assume a simple Gaussian distribution of nutrient concentrations, 
\begin{equation}
P(s) = f(s; \avUni,\sS^2) \ .
\end{equation}
Simple integration shows that the posterior distribution $P(s|s^*)$ is a Gaussian with mean 
\begin{equation}
E[ s | s^* ] = \f{\avUni}{1 + r^{-1}} + \f{s^*}{1+r} \ ,
\end{equation}
and variance $(\sM^{-2} + \sS^{-2})^{-1}$, for the dimensionless parameter $r \equiv \sM^2/\sS^2$, the ratio of variances of the conditional measurement distribution and the environmental nutrient distribution. Hence for the strictly convex cost function with $n > 1$, the optimal enzyme level is
\begin{equation}
\eOpt(s^*) = \left[ \f{s^*}{K \costConst \, n\left( 1 + r \right)} \right]^{\f{1}{n-1}} \ .
\end{equation}

\subsubsection*{Bimodal nutrient distribution}
We now assume an equiprobable mixture of two Gaussians, each with the same variance $\sS^2$:
\begin{align}
P(s) = \frac{1}{2}\left[f(s; \mu_{\rm L},\sS^2)+f(s; \mu_{\rm H},\sS^2)\right] \ .
\end{align} 
Here, $\mu_{\rm L}$ and $\mu_{\rm H}$ ($\mu_{\rm L} < \mu_{\rm H}$) are the mean levels of the nutrient $s$ in each environment. 
Making a change of variables to $\mu = (\mu_{\rm L}+\mu_{\rm H})/2$ and $\Delta\mu=\mu_{\rm H}-\mu_{\rm L}$, and evaluating the Gaussian integrals, the posterior for $s$ has a mean of
\begin{align}
E[ s | s^* ] = \f{s^*}{1+r} + \f{r}{1+r}\left[ \mu + \tfrac{1}{2}\Delta\mu \tanh \f{\Delta\mu(s^*-\mu)}{2(\sM^2 + \sS^2)} \right] \ .
\end{align}
Fig.~S1 shows optimal regulatory strategies as a function of $s^*$, across several values of $\sS$ and $\sM$.  

\subsubsection*{Multimodal nutrient distribution}
This model is easily extensible to several environmental modes.
\begin{subequations}
\begin{align}
P(s) &= \frac{1}{k}\sum_{i=1}^k f(s; \mu_i, \sS^2) \\
P(s^*) &= \f{1}{k} \sum_{i=1}^k f(s^*; \mu_i, \sM^2+\sS^2) \\
P(s|s^*) &= f(s; s^*, \sM^2) \f{\sum_{i=1}^k f(s; \mu_i, \sS^2) }{ \sum_{i=1}^k f(s^*; \mu_i, \sM^2+\sS^2) }  \\
E[s|s^*] &=  \sum_{i=1}^k \f{ f(s^*; \mu_i, \sM^2+\sS^2) }{ \sum_{j=1}^k f(s^*; \mu_j, \sM^2+\sS^2) } \f{\sM^2 \mu_i + \sS^2 s^*}{\sM^2+\sS^2}  \\
&= \f{1}{1+r} \, s^* + \f{r}{1+r} \sum_{i=1}^k \f{ f(s^*; \mu_i, \sM^2+\sS^2) }{ \sum_{j=1}^k f(s^*; \mu_j, \sM^2+\sS^2) } \mu_i \ .
\end{align}
\end{subequations}
In this case, the expectation is a weighted sum of terms, one for each Gaussian mode in the mixture.  The term corresponding to each mode $i$ is weighted by the likelihood that the measurement comes from that mode, $\exp\{-(s^*-\mu_i)^2/[2(\sM^2+\sS^2)]\}$.  Each term takes the form of a weighted sum of the mean $\mu_i$ of the $i$th Gaussian mode and the observation $s^*$, weighted by the uncertainties associated with the measurement ($\sM^2$) and with the distribution within a given Gaussian mode ($\sS^2$), respectively.

This model is also trivially generalized to an arbitrary prior over the different modes. For a prior probability $p_i$ that the environment is in Gaussian $i$ with distribution $f(s; \mu_i,\sS^2)$:
\begin{subequations}
\begin{align}
P(s) &= \sum_{i=1}^k p_i \, f(s; \mu_i, \sS^2) \\
P(s^*) &= \sum_{i=1}^k p_i \, f(s^*; \mu_i, \sM^2+\sS^2) \\
P(s|s^*) &= f(s^*; s,\sM^2) \f{\sum_{i=1}^k p_i\ f(s; \mu_i,\sS^2)}{\sum_{i=1}^k p_i\ f(s^*; \mu_i,\sS^2+\sM^2)}  \\
E[s|s^*] &= \sum_{i=1}^k  \f{ p_i \ f(s^*; \mu_i, \sM^2+\sS^2) }{ \sum_{j=1}^k p_j \ f(s^*; \mu_j, \sM^2+\sS^2) } \f{\sM^2 \mu_i + \sS^2 s^*}{\sM^2+\sS^2}  \\
&= \f{1}{1+r} \, s^* + \f{r}{1+r} \sum_{i=1}^k  \f{ p_i \ f(s^*; \mu_i, \sM^2+\sS^2) }{ \sum_{j=1}^k p_j \ f(s^*; \mu_j, \sM^2+\sS^2) }  \mu_i \ .
\end{align}
\end{subequations}

\subsection*{Dynamic environments and the value of memory}
Previously, we analyzed an environment where the nutrient signal was uncorrelated in time, so that $s_{\nT}$ and $s_{\nT-1}$ were statistically independent random variables, where $\nT$ indexes the nutrient signal in time. Now, we consider an environment with temporal structure. We ask how a cell can optimally combine measurements of a nutrient signal in time to optimally regulate the level of the enzyme: what regulatory strategy $\eOpt(s_{\nT}^*,s_{\nT-1}^*)$ maximizes the payoff $\payoff(e,s_{\nT})$. This task consists in choosing the enzyme level $e_{\nT}$ that, for given measurements $s_{\nT}^*$ and $s_{\nT-1}^*$, maximizes the expected payoff
\begin{subequations}
\begin{align}
\payoff(e_{\nT},s_{\nT}^*,&s_{\nT-1}^*) \equiv E[\payoff(e_{\nT},s_{\nT})|s_{\nT}^*,s_{\nT-1}^*] \\
&= \int \md s_{\nT} \ \payoff(e_{\nT},s_{\nT}) \, P(s_{\nT}|s_{\nT}^*,s_{\nT-1}^*) \ .
\end{align}
\end{subequations}

We proceed similarly to before, but now we derive the average value of the payoff function with respect to both past and current measurements. To this end, we derive an expression for $P(s_{\nT}|s_{\nT}^*,s_{\nT-1}^*)$ with two assumptions: first, that the environmental dynamics are Markovian, 
\begin{align}
P(s_{\nT},s_{\nT-1}) = P(s_{\nT}|s_{\nT-1}) \ P(s_{\nT-1}) \ ;
\end{align}
and secondly, that a measurement depends only on the current true nutrient  concentration via a time-invariant measurement distribution $P(s_{\nT}^*|s_{\nT})$: 
\begin{align}
P(s_{\nT}^*,s_{\nT-1}^*|s_{\nT},s_{\nT-1}) = P(s_{\nT}^*|s_{\nT})\ P(s_{\nT-1}^*|s_{\nT-1}) \ .
\end{align}

Given these assumptions,
\begin{subequations}
\begin{align}
P(&s_{\nT}|s_{\nT}^*,s_{\nT-1}^*) = \frac{P(s_{\nT}^*,s_{\nT-1}^*|s_{\nT}) P(s_{\nT})}{P(s_{\nT}^*,s_{\nT-1}^*)}  \\
&= \int \md s_{\nT-1} \, \frac{P(s_{\nT}^*,s_{\nT-1}^*|s_{\nT},s_{\nT-1})\ P(s_{\nT},s_{\nT-1})}{P(s_{\nT}^*,s_{\nT-1}^*)} \\
& = \int \ \md s_{\nT-1} \ P(s_{\nT}|s_{\nT-1}) P(s_{\nT-1}) \frac{P(s_{\nT}^*|s_{\nT})\ P(s_{\nT-1}^*|s_{\nT-1})}{P(s_{\nT}^*,s_{\nT-1}^*)} \ ,
\end{align}
\end{subequations}
where $P(s_{\nT+1}|s_{\nT})$ is the environmental transition probability and 
\begin{align}
P(s_{\nT}^*,s_{\nT-1}^*) =  \int \ \md s_{\nT} \ \md s_{\nT-1} \, P(s_{\nT}^*|s_{\nT})\ P(s_{\nT-1}^*|s_{\nT-1})\ P(s_{\nT}|s_{\nT-1}) P(s_{\nT-1}) \ .
\end{align} 
Thus the expected payoff is
\begin{align} 
E[\payoff(e_{\nT},s_{\nT})|s_{\nT}^*,s_{\nT-1}^*]  = \int \  \md s_{\nT} \ \md s_{\nT-1} \, \payoff(e_{\nT},s_{\nT}) \frac{P(s_{\nT}^*|s_{\nT}) P(s_{\nT-1}^*|s_{\nT-1})\ P(s_{\nT}|s_{\nT-1}) P(s_{\nT-1}) }{\int \ \md s'_{\nT} \ \md s'_{\nT-1} \, P(s_{\nT}^*|s'_{\nT}) P(s_{\nT-1}^*|s'_{\nT-1}) P(s'_{\nT}|s_{\nT-1}) P(s'_{\nT-1})} \ .
\end{align}
As previously, we also note that for our specific payoff function $\payoff(e_{\nT},s_{\nT}) = \f{e_{\nT} }{K} \, s_{\nT} - \costConst \, e_{\nT}^n$, the expected payoff, conditional on current and immediate past measurements, is 
\begin{subequations} 
\begin{align}
\payoff(e_{\nT},s_{\nT}^*,s_{\nT-1}^*) &= \int \md s_{\nT} \, \l(\f{e_{\nT}}{K} \, s_{\nT} - \costConst \, e_{\nT}^n\r) P(s_{\nT}|s_{\nT}^*,s_{\nT-1}^*) \\
&= \f{e_{\nT}}{K} \, E[ s_{\nT} | s_{\nT}^*,s_{\nT-1}^* ]- \costConst \, e_{\nT}^n \ .
\end{align} 
\end{subequations} 

We consider a mean-reverting environment with conditional distribution $P(s_{\nT}|s_{\nT-1}) = f(s_{\nT}; \avDyn + a [s_{\nT-1}-\avDyn], [1-a^2]\sS^2)$ that therefore has a constant marginal distribution $P(s_{\nT}) = f(s_{\nT}; \avDyn, \sS^2)$. The correlation of nutrient concentrations decays geometrically with $a$, 
\begin{equation}
\langle s_{\nT} \, s_{\nT+j} \rangle = a^n \sS^2 \ ,
\end{equation}
such that the correlation time, in units of discrete time steps, is
\begin{equation}
\tau_{\rm corr} \equiv \int \md j \ \langle s_{\nT} s_{\nT+j} \rangle = - \f{1}{\ln a} \ .
\end{equation}

As before, we assume a Gaussian measurement error $P(s_{\nT}^*|s_{\nT}) = f(s_{\nT}^*; s_{\nT}, \sM^2)$. Straightforward integration leads us to a relatively compact expression for the expected nutrient concentration given the current and previous measurements
\begin{align}
E[s_{\nT}|s_{\nT}^*,s_{\nT-1}^*] = \frac{[(1-a^2) + r]s_{\nT}^* + a\,r \, s_{\nT-1}^* + [(1-a)r + r^2] \avDyn }{(1-a^2) + 2 r + r^2} \ .
\end{align}
Fig.~S2 shows optimal regulatory strategies as a function of present and past readouts $s_{\nT}^*$ and $s_{\nT-1}^*$, across several values of $r$ and environmental persistence $a$.  

We can extend the expectation to depend on two past measurements in a derivation that is algebraically tedious but conceptually identical to the one above:
\begin{align}
&E[s_{\nT}|s_{\nT}^*,s_{\nT-1}^*,s_{\nT-2}^*]  = \\
&\frac{ [(1-a^2)^2 + (1-a^2)(2+a^2)r + r^2] s_{\nT}^* + [a(1-a^2)r + ar^2] s_{\nT-1}^* + a^2r^2 \, s_{\nT-2}^* + [(1-a)(1-a^2)r + (2-a-a^2)r^2 + r^3] \avDyn }{ (1-a^2)^2 + (3-2a^2+a^4)r + 3r^2 + r^3 } \notag
\end{align}

\section*{Acknowledgments}
We thank Hana El-Samad, Wendell Lim, Amir Mitchell, Michael Fischbach, and Hao Li for enlightening discussions, and especially Hyun Youk for detailed feedback on the manuscript.

\bibliography{OptimalStrategies}

\renewcommand{\thefigure}{S\arabic{figure}}
\setcounter{figure}{0}

\begin{figure}[!ht]
\begin{center}
\includegraphics[width=17.35cm]{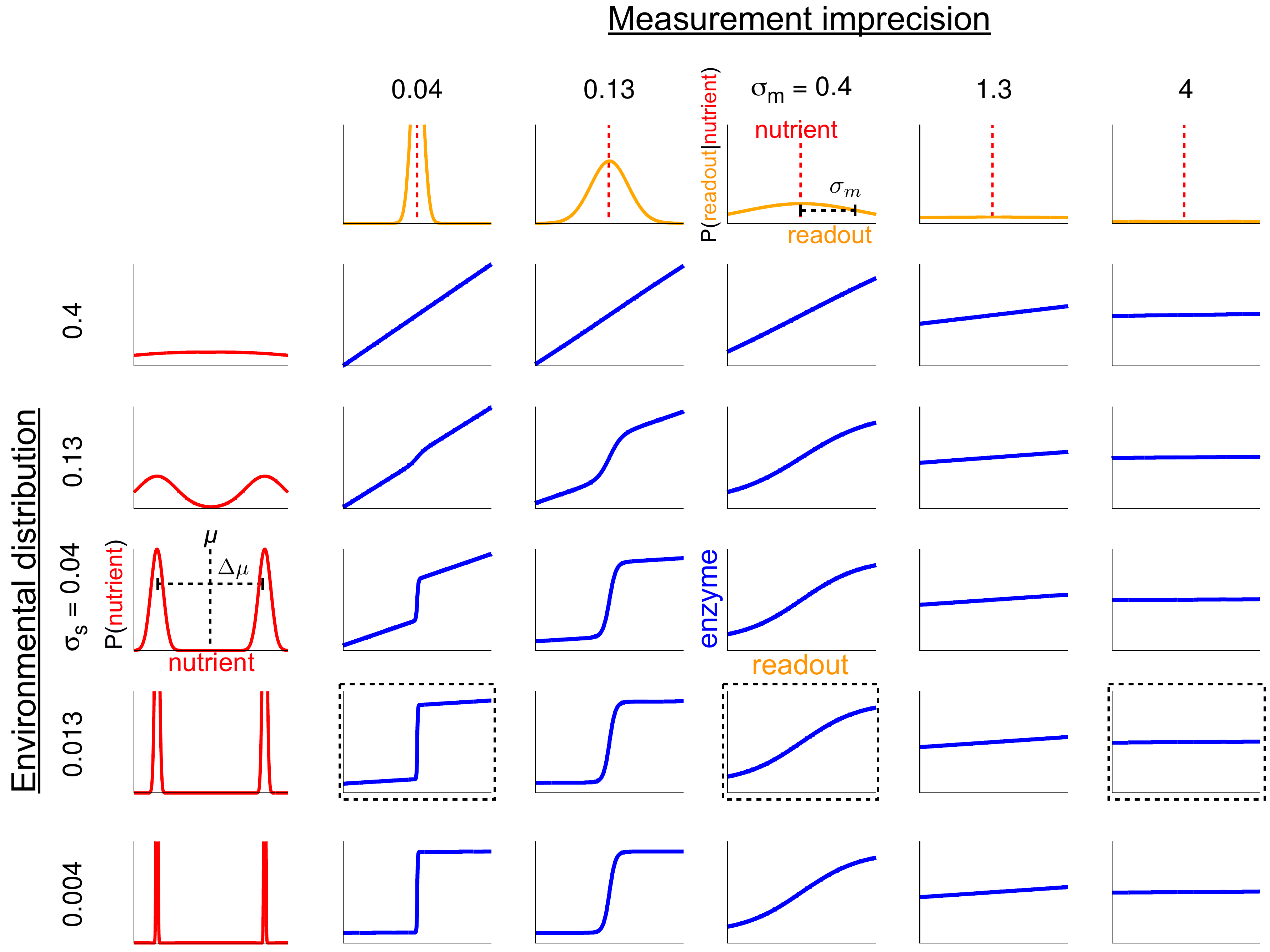}
\end{center}
\caption{{\bf Optimal regulatory strategy varies with environmental variability and measurement imprecision.} Blue curves plot optimal regulatory strategy as a function of cellular readout $s^*$, for bimodal environments of varying mode width (depicted in leftmost column) and for varying measurement imprecision (depicted in upper row). Black dashed boxes indicate the selected strategies shown in Fig.~\ref{fig:Fig4}.}
\label{fig:Supp1}
\end{figure}

\begin{figure}[!ht]
\begin{center}
\includegraphics[width=17.35cm]{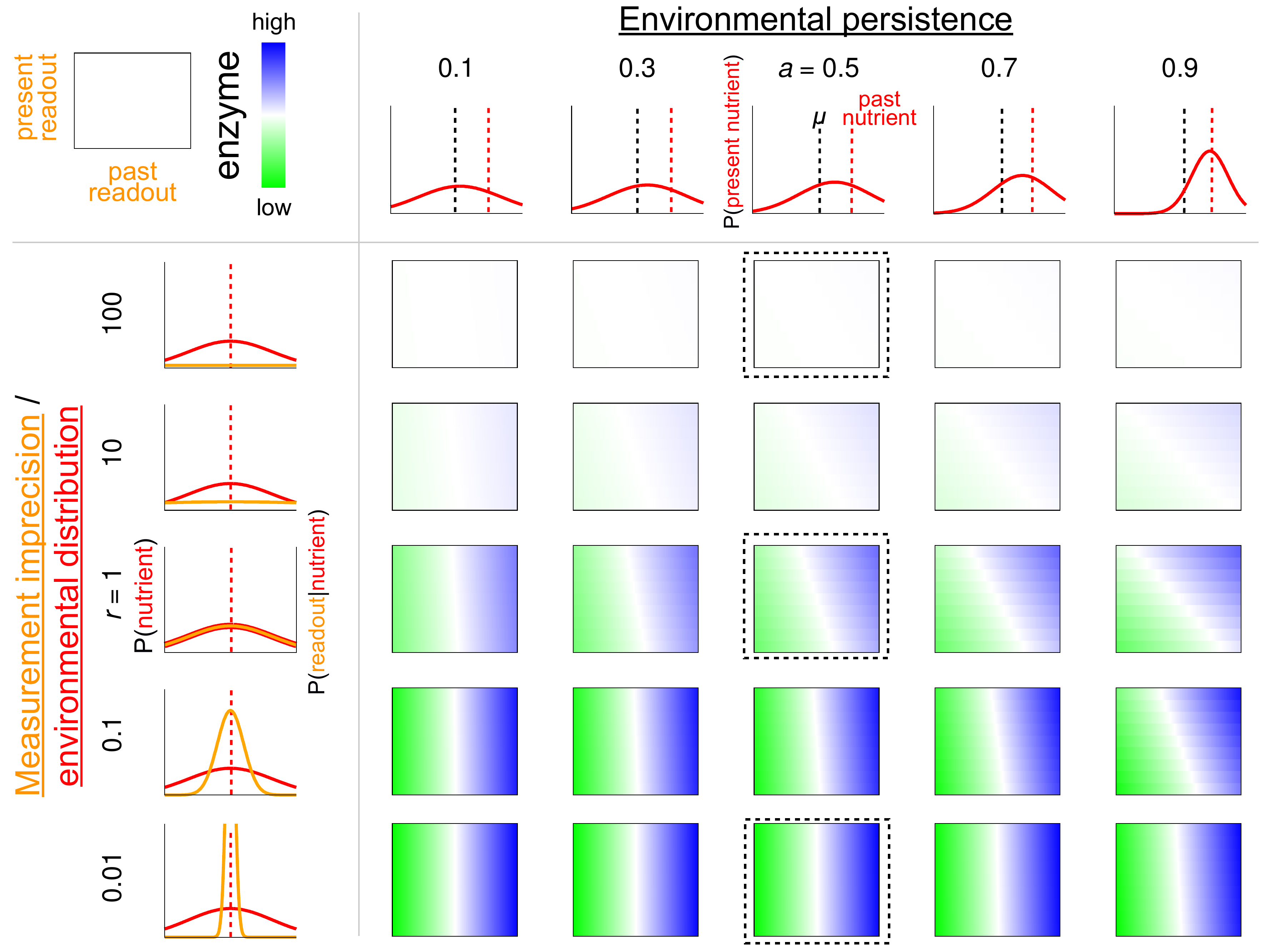}
\end{center}
\caption{{\bf Optimal regulatory strategy varies with environmental persistence and relative measurement imprecision.} Heat maps plot optimal regulatory strategy as a function of present readout $s^*_{\nT}$ ($x$-axis) and past readout $s^*_{\nT-1}$ ($y$-axis), for varying environmental variability and measurement precision (both depicted in leftmost column) and for varying environmental persistence. Environmental persistence is depicted in upper row as the probability distribution of present nutrient concentration, given steady-state mean $\mu$ (black dashed line) and previous nutrient concentration (red dashed line). Black dashed boxes indicate the selected strategies shown in Fig.~\ref{fig:Fig5}.}
\label{fig:Supp2}
\end{figure}

\end{document}